\documentstyle[12pt,psfig]{article}

\newcommand{\epsfig}{\psfig}
\newcommand{\p}{{\partial}}
\newcommand{\cF}{{\cal F}}
\newcommand{\cM}{{\cal M}}
\newcommand{\cG}{{\cal G}}
\newcommand{\hf}{\frac12}
\newcommand{\bea}{\begin{eqnarray}}
\newcommand{\eea}{\end{eqnarray}}
\newcommand{\be}{\begin{equation}}
\newcommand{\ee}{\end{equation}}
\newcommand{\ba}{\begin{array}}
\newcommand{\ea}{\end{array}}
\newcommand{\eps}{\epsilon}
\newcommand{\La}{\Lambda}
\newcommand{\lam}{\lambda}
\newcommand{\la}{\langle}
\newcommand{\ra}{\rangle}
\newcommand{\del}{\delta}
\newcommand{\Del}{\Delta}
\newcommand{\tla}{\tilde{\Lambda}_5}
\newcommand{\om}{\omega}
\newcommand{\Om}{\Omega}

\newcommand{\tr}{{\rm Tr}}
\newcommand{\cO}{{\cal O}}
\def\dfrac#1#2{ {\displaystyle\frac{#1}{#2}} } 
\def\ZZ{{\bf Z}}

\begin{document}

\begin{titlepage}
\thispagestyle{empty}
\vspace*{-0.568in}
\begin{flushright}
LMU-TPW  96-21\\
hep-th/9609xxx
\end{flushright}
\begin{center}
\bigskip\bigskip
\vspace*{2cm}
{\LARGE{ \bf Prepotentials in $N=2$ Supersymmetric $SU(3)$ \\ 
YM-Theory with Massless Hypermultiplets$^*$}}

\vskip2cm
Holger Ewen$^1$, Kristin F\"orger$^1$, Stefan Theisen$^{1,2}$
\vskip1cm
{\sl $^1$ Sektion Physik, Universit\"at M\"unchen \\
Theresienstr. 37, 80333 M\"unchen, Germany}
\vskip.5cm
{\sl $^2$ CERN, Theory Division, Geneva, Switzerland}
\end{center}
\vskip1in


\begin{abstract}
We explicitly determine the instanton corrections to the prepotential  
for $N=2$ supersymmetric $SU(3)$ Yang-Mills theory with massless 
hypermultiplets in the weak coupling regions $u\to\infty$ and $v\to\infty$.
We construct the Picard-Fuchs equations for $N_f<6$ 
and calculate the monodromies using Picard-Lefschetz theorem for $N_f=2,4$. 
For all $N_f<6$ the instanton corrections to the prepotential 
are determined using the relation between $\tr\la\phi^2\ra$ and the 
prepotential. 
\end{abstract}
\mbox{}
\vskip2cm
-----------------------------------\hfill\break
$^*$ {\small This work is partially
supported by GIF-the German-Israeli Foundation for Scientific Research, 
the DFG and by the European Commission TMR programme ERBFMRX-CT96-0045,
in which H.E, K.F. and S.T. are associated to HU-Berlin}
\end{titlepage}
\newpage
\setcounter{page}{1}

\section{Introduction}

In the last two years duality has become a very important tool in 
supersymmetric gauge theories as well as in string theory.
The basic idea was developed by Seiberg and Witten \cite{sw}
(for reviews, see e.g. Bilal \cite{bilal} or Lerche \cite{lerche})
who found, at the two-derivative level, the exact non-perturbative
low energy Wilsonian effective action of $N=2$ supersymmetric gauge theory
with gauge group $SU(2)$ by using duality and the selfconsistent 
assumption of massless monopoles and dyons in the strong coupling 
region of the moduli space ${\cal M}$.

The main technical point in \cite{sw} is that the moduli space ${\cal M}$ 
coincides with the moduli space of an auxiliary elliptic curve\footnote{The 
appearance of Riemann surfaces in the solution of $N=2$ supersymmetric YM theory finds an 'explanation'
in the string theory context \cite{klvk,klvw}.}.
Subsequently, generalizations 
to gauge groups $SU(N_c)$ without \cite{klyt,klt,af} and with matter (in the 
fundamental representation) \cite{dsh,ho,iy,oh} have been worked out.
Extensions to other groups, $SO(N_c)$ and $Sp(N_c)$  
\cite{ds,bl}  as well as to exceptional groups \cite{lw,ds2} are also known.

The field content of $N=2$ gauge theories for arbitrary gauge group $G$ 
consists of an $N=2$ chiral multiplet which contains
a vector field, two Weyl fermions and a complex 
scalar, all in the adjoint representation of the gauge group. 
In addition to the gauge sector we can have 
$N_f$ hypermultiplets, each  containing two Weyl fermions 
and two complex bosons, transforming in some representation of $G$. 
The theory has flat directions with nonvanishing expectation
values for ${\rm Tr}\phi^2$, along which the gauge group is 
generically broken to the Cartan
subalgebra, e.g. $SU(3)$ is broken to $U(1)^{\otimes 2}$.

The Wilsonian low energy effective action is specified 
by a single holomorphic 
prepotential $\cF$ and can be expressed in terms of $N=1$ chiral 
multiplets $A_i$, whose scalar component we denote by $a_i$, 
and $N=1$ $U(1)$ gauge multiplets $W_{\alpha i}$ ($i=1,\ldots,N_c-1$):
\be
{\cal L}_{\rm eff}=\frac{1}{4\pi}{\rm Im}\bigg\{\int d^4\theta \;
\p_i{\cal F}(A)\,\bar{A}^i+\frac{1}{2}\int d^2\theta\;
\p_i\p_j{\cal F}(A)\; W_{\alpha}^i W^{\alpha j}\bigg\}
\ee
where $\p_i\cF=\dfrac{\p\cF}{\p A^i}$, and 
\be
\p_i\p_j\cF(a)=\tau_{ij} (a)=\left(\frac{8\pi i}{g^2}+
\frac{\theta}{\pi}\right)_{ij}
\ee
is the field dependent coupling constant. The metric on the 
quantum moduli space $\cM$, $ds^2={\rm Im}(da_{Di}d\bar a_i)$
where $a_{Di}=\frac{\p\cF}{\p a_i}$ is the magnetic dual of $a_i$, 
has singularities at which the local effective action breaks down due to 
certain BPS states becoming massless. 
Loops in moduli space around these singularities yield monodromies 
of the section $\Pi=\left(\vec{a}_D\atop\vec{a}\right)$.
The global monodromy properties fix the bundle. 
Once we know $a_D(a)$, the prepotential can be 
obtained by integration. To obtain $a$ and $a_D$ we use the fact that they are
period integrals of a particular meromorphic differential $\lambda$ on 
a hyperelliptic curve whose period matrix is $\tau$. 
In the weak coupling region the prepotential is generally 
given by $\cF=\cF_{\rm class}+\cF_{\rm 1-loop}+\cF_{\rm inst}$ \cite{ds2}:
\bea\label{1lp}
\cF_{\rm class}&=&\frac{\tau_{\rm class}}{2}\sum_{\alpha\in\Del_+} 
\la\alpha,a\ra^2\nonumber\\
\cF_{\rm 1-loop}&=&\frac{i}{4\pi}\sum_{ \alpha\in\Del_+}
\la\alpha,a\ra^2 \ln\Big(\frac{\la\alpha,a\ra^2}{\La_{N_f}^2}\Big)
-N_f\frac{i}{8\pi}\sum_{{w}}\langle{w},a\rangle^2
\ln\Big(\frac{\la{w},a\ra^2}{\La_{N_f}^2}\Big)
\\
\cF_{\rm inst}&=&\sum_{n=1}^\infty \cF_n(a)\La_{N_f}^{(2 N_c-N_f) n}\nonumber
\eea
The sums are over the positive roots $\Del_+$ of $G$ and the weights 
of the representation of the hypermultiplets, respectively.

The exact results obtained from duality predict the precise form of the 
instanton corrections to the effective action. 
In the weak coupling regime various checks of these results have been 
performed by explicit instanton calculations in the microscopic theory 
at the one and two instanton level \cite{ahsw,hs,is,dkm,ft,fp}.

In this paper we compute the prepotential for $N=2$ supersymmetric $SU(3)$ 
gauge theory with $N_f<6$ flavors in the fundamental representation. 
Although both the perturbative and the 
nonperturbative instanton corrections can in principle be obtained 
by explicitly doing the period integrals, 
we are attacking this problem via the Picard-Fuchs 
equations for the periods of the appropriate hyperelliptic curves, following \cite{klt}.
The period integrals then only have to be solved to leading orders 
to determine which linear combinations of the given system of solutions 
of the Picard-Fuchs equations correspond to $(a_{Di},a_i)$. 
Given the periods in one patch of ${\cal M}$ we then get them everywhere
in ${\cal M}$ by analytic continuation. It is however often easier to solve the
Picard-Fuchs equations in various patches and again adjust coefficients
by computing the period integrals to leading orders. 
The monodromy can then be read off from the periods and, as a check, 
shown to coincide up to conjugation with the monodromies obtained from the 
Picard-Lefschetz formula \cite{klt}.
Using the methods outlined above,  
we explicitly compute the periods, the monodromies and  
the prepotential  $\cF$ for 
$N_f=2$ and $N_f=4$. The one and two instanton contributions will 
be given explicitly.
Our analysis will, however, be restricted to the weak coupling regime.
We then show, how the instanton corrections for the prepotential can alternatively be 
obtained from the  Picard-Fuchs operators and the relation between 
$\tr\la\phi^2\ra$ and $\cF$, derived in \cite{m,sty}.
We do this explicitly for all $N_f<6$.

While this paper was being proofread, a paper \cite{dkp} was posted on hep-th  
which treats the same problem by explicitly doing the period integrals.

\section{Picard-Fuchs Operators for $A_2$ with $N_f<6$}

We start from the hyperelliptic curves associated with the gauge group $A_2$ 
with matter in the form given in \cite{mn}. 
We restrict ourselves here to the case of massless matter where the 
curves for $N_f< 6$ are given by:
\be\label{cur}
y^2=W(x;u,v)^2-F(x;\La_{N_f})
\ee
Here $W(x;u,v)=x^3-u x-v$ is the $A_2$-type simple singularity with the identification
\bea
u&=&{\rm Tr}\langle \phi^2\rangle\nonumber\\ 
v&=&{\rm Tr}\langle \phi^3\rangle
\eea
as the gauge invariant coordinates on the moduli space ${\cal M}$ 
\cite{klt,klyt}.
Under the anomaly free global subgroup $\ZZ_{4N_c-2N_f} \subset U(1)_R$
they have $R$ charge $4$ and $6$ respectively. $F$ is given by
\be
F(x;\La_{N_f})=\Lambda_{N_f}^{6-N_f} (x-\del_{N_f,5}\frac{\La_5}{12})^{N_f} 
\ee
$\Lambda_{N_f}$ is the dynamically generated scale of the theory, which 
can be matched to the scale of the microscopic theory \cite{fp}. 
In the limit $\Lambda_{N_f}\to 0$ the curves 
corresponding to the classical moduli space are recovered.

The singularities of the quantum moduli space are the zero locus 
of the discriminant of the curve. The discriminants are:
\bea
\Del_{N_f=1}&=&\La_1^{15}\Big(-3125\; \La_1^{15}+256\;
\La_1^5\;u^5+22500\;\La_1^{10} \;u \;v-1024\; u^6\; v
-43200\;\La_1^5 u^2 v^2\nonumber\\
& & +13824 \;u^3 \;v^3-46656\; v^5\Big)\\
\Del_{N_f=2}&=&\La_2^{12}\;v^2\Big(-4\;\La_2^6+12\;\La_2^4\;u-12\;
\La_2^2\;u^2+4\;u^3-27\;v^2\Big)
\Big(-4\;\La_2^6-12\;\La_2^4\; u\nonumber\\
& & -12\;\La_2^2\;u^2-4\;u^3+27\;v^2\Big)\label{disc2}\\
\Del_{N_f=3}&=&\La_3^9\;v^3\Big(-108\;\La_3^6\;u^3-1024\;u^6-729\;\La_3^9\;
v-8640\;\La_3^3\;u^3\;v-8748\;\La_3^6\;v^2\nonumber\\
& & +13824\; u^3\;v^2
-34992\;\La_3^3\;v^3-46656\; v^4\Big)\\
\Del_{N_f=4}&=&\La_4^6\; v^4\Big(\La_4^2\;u^2+4\; u^3+4\;\La_4^3\; 
v+18\;\La_4 \;u \;v-27\; v^2\Big)\Big(-\La_4^2\;u^2-4\; u^3+
4\;\La_4^3\; v\nonumber\\
& & +18\;\La_4 \;u \;v+27\; v^2\Big)\label{disc4}\\
\Del_{N_f=5}&=&\tla^3\Big(-\tla^3+\tla \;u+v\Big)^5\Big(-20601\tla^{12}
+67635\tla^{10}\;u-78840\tla^8\;u^2
\nonumber\\
& & +36288\;\tla^6\;u^3-5895\;\tla^4\;u^4+117\;\tla^2\;u^5-16\;u^6+62775\;
\tla^9\;v-143775\;\tla^7\;u\; v\nonumber\\
& & +90045\;\tla^5\;u^2\;v-17145\;\tla^3\;u^3\; v+36\;\tla \;u^4\; 
v-68121\;\tla^6\; v^2+71280\;\tla^4\; u \;v^2\nonumber\\
& & -18495\;\tla^2\; u^2\; v^2+216 \;u^3\; v^2+12825\;\tla^3\; 
v^3-6075\;\tla \;u \;v^3-729 \;v^4\Big)
\eea
where $\tla=\frac{\La_5}{12}$. 
Obviously the moduli space for odd number of flavors is much more complicated than
the one for even flavors where the discriminant factorizes.

The meromorphic differential on the Riemann surface of genus $2$ 
corresponding to the hyperelliptic 
curve (\ref{cur}) for $N_f<6$ is \cite{aps}:
\bea\label{la}
\lam&=&\frac{x}{2\pi i y F}\Big(2FW'-WF'\Big)\nonumber\\
&=&\frac{x}{2\pi i y (x-\frac{\La_5}{12}\del_{N_f,5})}
\Big\{(6-N_f) x^3+(N_f-2) u x+N_f v-
\del_{N_f,5}\frac{\lam_5}{6}(3 x^2-u)\Big\}
\eea

The components of the section $\Pi$ are given as integrals over 
the meromorphic differential
\be
a_i=\int_{\alpha_i} \lam \hspace{1cm}{\rm and}\hspace{1cm} 
a_{Di}=\int_{\beta_i}\lam
\ee 
where  $\alpha_i$ and $\beta_i$ are a symplectic basis of homology 1-cycles 
on the curves (\ref{cur}), i.e. $\alpha_i\cap\beta_j=-\beta_j\cap\alpha_i=\del_{ij}$ 
and $\alpha_i\cap\alpha_j=\beta_i\cap\beta_j=0$.

The Picard-Fuchs operators constitute a system of partial differential 
operators of second order for the periods of a holomorphic differential.
A basis of holomorphic differentials on the $A_2$-curves is 
$\p_u\lam$, $\p_v\lam$.
We now proceed similar to \cite{klt} by considering first and second derivatives 
of $\p_v\lam$ with respect to $u$ and $v$.
This produces expressions of the form $\frac{\phi(x)}{y^n}$ where $\phi(x)$ are 
polynomials in $x$, $u$ and $v$.
The power of $y$ in the denominator as well as the degree of the 
polynomials in the numerator 
can be reduced by various identities to one and four, respectively.
Between these one can find two nontrivial linear combinations which 
vanish up to total derivatives.
They give rise to the second order Picard-Fuchs equations 
$\tilde{\cal L}_i\p_v\Pi=0$, with 
$\tilde{\cal L}$ of the general form:
\bea
\tilde{\cal L}_{(1)}&=&c_1^{(1)}\p^2_{uu}+c_2^{(1)}\p_{uv}^2+
c_3^{(1)}\p_u+c_4^{(1)}\p_v+c_5^{(1)}\\
\tilde{\cal L}_{(2)}&=&c_1^{(2)}\p^2_{vv}+c_2^{(2)}\p_{uv}^2+
c_3^{(2)}\p_u+c_4^{(2)}\p_v+c_5^{(2)}
\eea
where the coefficients $c_j^{(i)}$ are polynomials in $u$ and $v$.

For the reduction procedure one uses an identity derived from 
the fact that the discriminant 
can be written in the form $\Del=a(x) y^2+b(x) (y^2)'$ where 
$a(x)=\sum_{i=0}^4 a_ix^i$ and $b(x)=\sum_{i=0}^5b_ix^i$.
The two identities (up to total derivatives) which were used are: \\
(i) the power of $1/y$ is reduced by two through
\be
\frac{\phi(x)}{y^n}=\frac{1}{\Delta y^{n-2}}
\Big\{a\phi+\frac{2}{n-2}(b \phi')\Big\}
\ee
(ii) the powers of $x$ in the numerator can be reduced using  
\be
\frac{x^k}{y^l}=-\frac{x^{k-n-2}}{y^l(2-n l+2 k)}\Big((2-l) x 
\varphi+2 (k-n+1)\psi\Big)
\ee
for $k\neq (n l -2) / 2$, where $\varphi$ and $\psi$ are defined 
via $y^2=x^n+\psi(x)$ and
$(y^2)'=n x^{n-1}+\varphi(x)$ and $n=6$ for the $SU(3)$ case.

Since we are treating massless matter, $(\vec{a}_D,\vec{a})$ 
transforms irreducibly under monodromy.
It is therefore possible to find Picard-Fuchs operators 
 $\cal L$ with $\p_v{\cal L}_i\Pi=\tilde{\cal L}_i\p_v\Pi=0$.
If we pull $\p_v$ through 
$\tilde{\cal L}_i$ we obtain the following set of Picard-Fuchs operators:\footnote{Picard-Fuchs 
operators for the curves given in \cite{ho} were
derived in \cite{iy}.}
\bea
\lefteqn{N_f=1\;:}\nonumber\\
{\cal L}_{(1)}&=&16\Big(25\;\La_1^5 u^2-84\; u^3 v-405\; v^3\Big)\p^2_{uu}
+\Big(-625\;\La_1^{10}+3300\;\La_1^5\; u\; v\nonumber\\
& & -3456 \;u^2 \;v^2\Big)\p^2_{uv}+12 v\Big(25\;\La_1^5-36\; u\; 
v\Big)\p_v+4 \Big(25\;\La_1^5-84\; u\; v\Big)\label{pf11}\\
{\cal L}_{(2)}&=&4 \Big(25\;\La_1^5\; u^2-84 \;u^3\; v-405 \;
v^3\Big)\p^2_{vv}+\Big(1125\;\La_1^5\; v-64 \;u^4-2160\; u\;
 v^2\Big)\p^2_{uv}\nonumber\\
& & +4\Big(-16\; u^3-135\; v^2\Big)\p_v-180 \;v\label{pf12}\\[2ex]
\lefteqn{N_f=2\;:}\nonumber\\
{\cal L}_{(1)}&=&\Big(-8\;\La_2^4 \;u+8 \;u^3+27 \;v^2\Big)\p^2_{uu}
+6 v\Big(\La_2^4+3\;u^2\Big)\p^2_{uv}+2\; u\label{pf21}\\
{\cal L}_{(2)}&=&3 v\Big(8\; u^3+27 \;v^2-8 \;\La_2^4\; u\Big)\p^2_{vv}
+4 \Big(2(\La_2^4-u^2)^2+27\; u\; v^2\Big)\p^2_{uv}\nonumber\\
& &+ \Big(8\; u^3+27\; v^2-8\;\La_2^4 \;u\Big)\p_v+9 \;v\label{pf22}\\[2ex]
\lefteqn{N_f=3\;:}\nonumber\\
{\cal L}_{(1)}&=&\Big(9\;\La_3^6+72\;\La_3^3 \;v+64 \;u^3+144\; 
v^2\Big)\p^2_{uu}+4 \;u^2\Big(-\La_3^3+32 \;v\Big)\p^2_{uv}\nonumber\\
& & +4 u\Big(-\La_3^3-4 \;v\Big)\p_v+16\; u\label{pf31}\\
{\cal L}_{(2)}&=&9 v\Big(9\;\La_3^6+72\;\La_3^3 v+64 \;u^3+144 \;
v^2\Big)\p^2_v+u\Big(27\;\La_3^6+540\;\La_3^3 v+256\; u^3+\nonumber\\
& & 1728\; v^2\Big)\p^2_{uv}+\Big(27\;\La_3^6+216\;\La_3^3 v+256\; 
u^3+432 \;v^2\Big)\p_v+36\Big(\La_3^3+4\; v\Big)\label{pf32}\\[2ex]
\lefteqn{N_f=4\;:}\nonumber\\
{\cal L}_{(1)}&=&\Big(4\;\La_4^4 \;u+31\La_4^2\; u^2+60 \;u^3+81\; 
v^2\Big)\p^2_{uu}+4 v\Big(2\;\La_4^4+15\;\La_4^2 \;u+27 u^2\Big)
\p^2_{uv}\nonumber\\
& & +3 v\Big(-2\;\La_4^2-9\; u\Big)\p_v+4\;\La_4^2+15 \;u\label{pf41}\\
{\cal L}_{(2)}&=&v\Big(4\;\La_4^4 u+31\;\La_4^2 u^2+60 \;u^3
+81 v^2\Big)\p^2_{vv}+2\Big(\La_4^4 u^2+8\; \La_4^2u^3+9\;
\La_4^2 v^2\nonumber\\
& & +16 \;u^4+54\; u v^2\Big)\p^2_{uv}+\Big(2\;\La_4^4 u+16\;
\La_4^2u^2+32\;u^3+27 v^2\Big)\p_v+9 v\label{pf42}\\[2ex]
\lefteqn{N_f=5\;:}\nonumber\\
{\cal L}_{(1)}&=&\Big(1152\;\tla^9+1908\;\tla^7 u+6228\;
\tla^6 v-1860\;\tla^5 u^2-4680\;\tla^4 u v+492\;\tla^3 u^3+\nonumber\\
& & 396\;\tla^3 v^2+1140\;\tla^2 u^2 v-52\;\tla u^4-306\;\tla\; 
u v^2-132\; u^3 v-81\; v^3\Big)\p^2_{uu}+\nonumber\\
& & \Big(-45\;\tla^{10}-2070\;\tla^8 u-3960\;\tla^7 v+150\;
\tla^6 u^2-1665\;\tla^5 u v+354\;\tla^4 u^3-3960\;\tla^4 v^2\nonumber\\
& & +1866\;\tla^3u^2v-29\;\tla^2 u^4 +2520\;\tla^2 u v^2-165\;
\tla u^3 v+270\;\tla v^3-216\; u^2 v^2\Big)\p^2_{uv}+\nonumber\\& & 
\Big(-630\;\tla^8-480\;\tla^6 u-1755\;\tla^5 v+306\;\tla^4 u^2+714\;
\tla^3 u v-16\;\tla^2 u^3-180\;\tla^2 v^2-\nonumber\\
& & 15\;\tla u^2 v+81\; u v^2\Big)\p_v +15\;\tla^5+48\;\tla^3 u+120\;
\tla^2 v-13\;\tla u^2-33\; u v\label{pf51}\\
{\cal L}_{(2)}&=&\Big(1152\;\tla^9+1908\;\tla^7 u+6228\;\tla^6 v-1860\;
\tla^5 u^2-4680\;\tla^4 u v+492\;\tla^3 u^3+396\;\tla^3 v^2\nonumber\\
& & +1140\;\tla^2 u^2 v-52\;\tla u^4-306\;\tla\; u v^2-132\; u^3 v-81\; 
v^3\Big)\p^2_{vv}+\Big(-4770\;\tla^8+\nonumber\\
& & 4860\;\tla^6 u -3285\;\tla^5 v-2298\;\tla^4 u^2+2538\;
\tla^3 u v+648\;\tla^2 u^3+360\;\tla^2 v^2-\nonumber\\
& & 465\;\tla u^2 v- 80\; u^4-108 \;u v^2\Big)\p^2_{uv}+
\Big(2340\;\tla^6-2112\;\tla^4 u+252\;\tla^3 v+\nonumber\\
& & 672\;\tla^2 u^2-120\;\tla u v-80\; u^3-27 \;v^2\Big)\p_v
+3\Big(18\;\tla^3-8\;\tla u-3\; v\Big)\label{pf52}
\eea

For the pure YM case $N_f=0$ the Picard-Fuchs equations were determined in \cite{klt} 
where it was found that they form an Appell system of type $F_4$.
For $N_f\neq 0$ we could not identify the system of Picard-Fuchs operators with any 
of the generalized hypergeometric systems discussed in the mathematical literature.

\section{$N_f=4$} 
\subsection{Monodromies}

The singularities of the moduli space are at the zero loci of the 
discriminant $\Del_{N_f}=0$. Here two or more roots of the curve 
$y^2=\prod_{i=1}^6(x-e_i)$ coincide and the associated genus two 
Riemann surface becomes singular.  
That is, some homology cycle 
$\nu=\vec{q}\cdot\vec{\alpha}+\vec{g}\cdot\vec{\beta}$ 
with magnetic charge vector 
 $\vec{g}=(g_1,g_2)^T$ and electric charge vector 
$\vec{q}=(q_1,q_2)^T$ vanishes, indicating that dyons with charge vector 
 $\vec{\nu}=(\vec{g},\vec{q})$ become massless.
Monodromies around these singularities have this charge 
vector as their left eigenvector
with eigenvalue one: $\vec{\nu}M_{\nu}=\vec{\nu}$. 
After the choice of a fixed base point in the moduli space, 
the monodromies from loops around various 
 branches $\Delta_{N_f}=0$ can be calculated from the Picard-Lefshetz formula 
\cite{klt}, which says that the monodromy action for a given cycle $\gamma$ 
is determined by the vanishing cycle $\nu$ of the singularity as 
 $M_{\nu}: \gamma\to\gamma-(\gamma\cap\nu)\nu$. 
This can easily be calculated after decomposing the vanishing 
cycles into the homology basis
 $\alpha_i$, $\beta_i$. This gives the monodromy in matrix form:
\be\label{pl}
M_{(\vec{g},\vec{q})}=\left(\ba{cc}{\bf 1}
+\vec{q}\otimes\vec{g}&\vec{q}\otimes\vec{q}\\\-
\vec{g}\otimes\vec{g}&{\bf 1}-\vec{g}
\otimes\vec{q}\ea\right)
\ee

We fix the homology basis as in figure 1:

\begin{figure}[htp]\label{cyc}
\begin{center}
\input{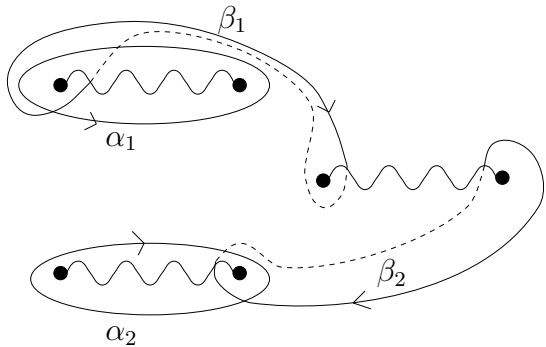}
\end{center}
\caption{Basic cycles $\alpha_i$ and $\beta_i$ in the $x$ plane}
\end{figure}

In the $v=1$ plane in the moduli space ${\cal M}_{N_f=4}$ we fix 
the reference point $u=-2$ and move to the six singular branches (see below),
tracing the motion of the roots in the $x$ plane.  
This results in the vanishing cycles shown in figure 2. They are
\bea
\nu_1&=&(1,0,0,0)\nonumber\\
\nu_2&=&(0,1,-1,2)\nonumber\\
\nu_3&=&(1,0,2,-1)\nonumber\\
\nu_4&=&(0,1,0,0)\\
\nu_5&=&(1,1,1,0)\nonumber\\
\nu_6&=&(1,1,0,-1)\nonumber
\eea
from which one can easily calculate the monodromy using (\ref{pl}).

\begin{figure}[htp]
\begin{center}
\input{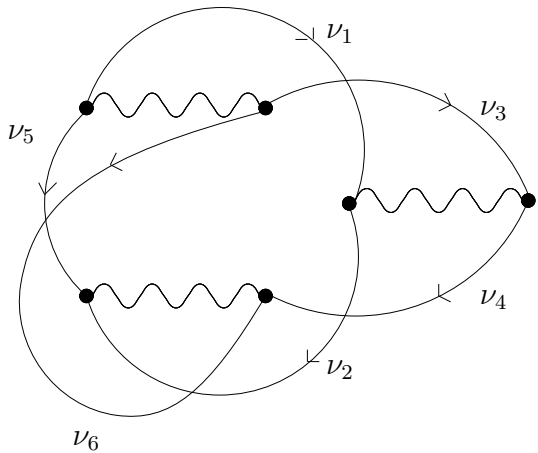}
\end{center}
\caption{Vanishing cycles for $N_f=4$}
\end{figure}

In figure 3 we give the structure of the moduli space ${\cal M}_{N_f=4}$.
The figure shows the zeroes of the discriminant (\ref{disc4}) 
with $\La_4=1$ and real $v$, projected onto the plane Im$u=0$.
There appear cusps at the points $(u,v)=(0,0)$ and 
$(-\frac{1}{3},\pm\frac{1}{27})$ and nodes 
at $(u,v)=(-\frac{1}{4},0)$ and $(-\frac{2}{9},\pm\frac{2}{81\sqrt{3}})$.
The four branches extending to the right of the vertical axis are real.
The shown branches which extend to Re~$u \to -\infty$ in fact each represent two 
branches with opposite Im~$u\neq 0$.
At the nodes the vanishing cycles do not intersect, that is  
$\nu_i\cap\nu_j=0$ 
and the corresponding monodromies commute $[M_ {\nu_i},M_{\nu_j}]=0$, 
in accordance with the van~Kampen relations.
This is a necessary condition for two dyons to condense 
simultaneously. They are then mutually local and 
can be described by a local effective action.

\begin{figure}
\begin{center}
\psfig{figure=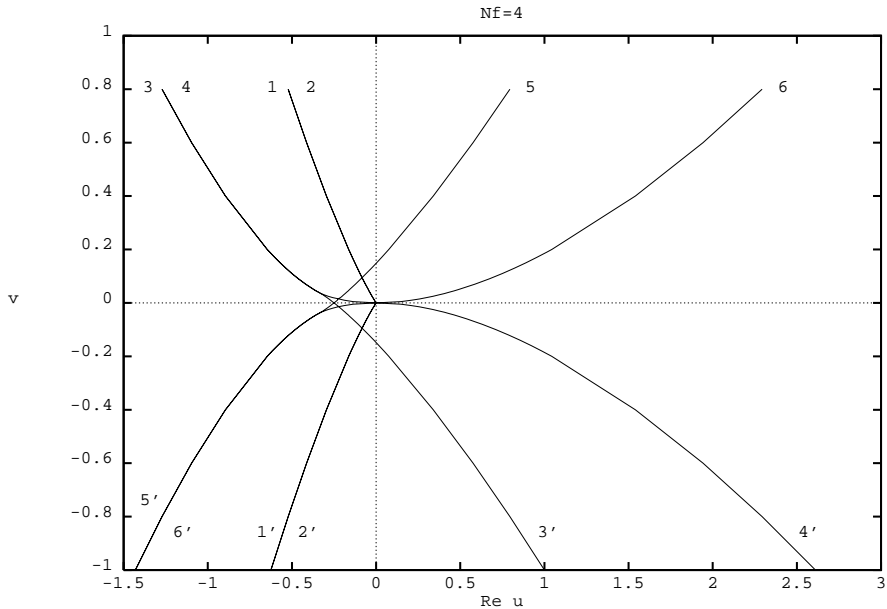,height=3in,angle=270}
\end{center}
\caption{Moduli space for $N_f=4$}
\end{figure}

We look at two weak-coupling regions of the moduli space: 
one has large $u$ for $v={\rm const.}$, the other has large $v$ with $u={\rm const.}$.
Making a loop in the first asymptotic region, we encircle for generic $v$ six lines,
in the second case at $u>0$ four lines.
The monodromies for these regions correspond to
going around all singular loci in the chosen plane, 
starting from the fixed base point. The semiclassical monodromy in 
the regions $u\to\infty$ and $v\to\infty$ are given by:
\bea
M_{u\to\infty}&=&r_2\cdot r_3\cdot r_1=
\left(\ba{rrrr}0&-1&-5&6\\-1&0&8&-5\\0&0&0&-1\\0&0&-1&0\ea\right)
\label{muinfty}\\
M_{v\to\infty}&=&(r_3\cdot M'_4\cdot M'_3)^{-1}=
\left(\ba{rrrr}-1&-1&-3&0\\1&0&6&-3\\0&0&0&-1\\0&0&1&-1\ea\right)
\label{mvinfty}
\eea
where $r_1$, $r_2$, $r_3$ are the classical Weyl group generators:
\bea
r_1 & = & M_3 M_1=\left(\ba{rrrr} 
{-1} & 0 & {4} &- 2 \\ 
1 & 1 & -2 & {1}\\ 
0 & 0 & {-1} & 1\\ 
0 & 0 & 0 & 1 \ea\right)\\
r_2 & = & M_2 M_4=\left(\ba{rrrr} 
1 & 1 & 1 &- 2\\ 
0 & -1 &- 2 & 4\\ 
0 & 0 & 1 & 0\\ 
0 & 0 & 1 & -1\ea\right)\\
r_3 & = & M_5 M_6=\left(\ba{rrrr} 
0 & -1 & 1& 2\\
 -1 & 0 & 0 & 1\\
 0 & 0 & 0 & -1\\
 0 & 0 & -1 & 0\ea\right)
\eea
and $M'_3$ and $M'_4$ are the monodromies associated 
to the vanishing cycles $\nu'_3=(0,1,0,-2)$ and $\nu'_4=(0,1,-1,0)$, 
where $M'_i$ denote monodromies associated with 
branches with Re~$v<0$.

\subsection{Solution in the semiclassical regions}

We are now going to compute the period integrals in the two semiclassical 
regions $v\to\infty$ and $u\to\infty$. 
In each of these regions we find a basis for the solutions of the Picard-Fuchs equations
consisting of two power series and two logarithmic solutions.
To match them with the four periods we analytically compute the period integrals
to leading orders.

For $v\to \infty$ we make the power series ansatz:
\be
\omega=\sum_{n\ge 0, m\le 0}c(n,m)u^{n+s}v^{m+t}
\ee
One finds $(s,t)=(0,\pm\frac{1}{3})$. 
For each set of indices one can find one power series solution and one
logarithmic solution of the form:
\be
\Omega=\om \ln\frac{u^a}{v^b} + \sum_{n\ge 0, m\le 0}d(n,m)u^{n+s}v^{m+t}
\ee
$a$ and $b$ are constants to be determined.
As we did not find a solution in closed form of the recursion relations for $c(n,m)$ and $d(n,m)$,
we give here only the first few terms of the solutions.

For $(s,t)=(0,\frac{1}{3})$ we get the following expansion:
\bea
\omega_1&=&v^{1/3}-\frac{35}{104976}\frac{\La_4^6}{v^{5/3}} 
-\frac{7}{1944}\frac{u}{v^{5/3}}\La_4^4-
\frac{19019}{11019960576}\frac{\La_4^{12}}{v^{11/3}}
-\frac{\La_4^2}{81}\frac{u^2}{v^{5/3}}
-\frac{1}{81}\frac{u^3}{v^{5/3}}+\dots\\
\Omega_1&=&\omega_1\ln\frac{\La_4^3}{v}-\frac{137}{209952}
\frac{1}{v^{5/3}}\La_4^6-
\frac{11}{1296}\frac{u}{v^{5/3}}\La_4^4-\frac{39737}{9183300480}
\frac{1}{v^{11/3}}\La_4^{12}\nonumber\\
& & \hbox{}-\frac{\La_4^2}{27}\frac{u^2}{v^{5/3}}
-\frac{47279}{510183360}\frac{u}{v^{11/3}}\La_4^{10}-
\frac{1}{18}\frac{u^3}{v^{5/3}}+\ldots
\eea
and for the second set of indices $(s,t)=(0,
-\frac{1}{3})$:
\bea
\omega_2&=&\frac{\La_4^2}{v^{1/3}}+6\frac{u}{v^{1/3}}
+\frac{385}{419904}\frac{\La_4^8}{v^{7/3}}+
\frac{55}{4374}\frac{u}{v^{7/3}}\La_4^6+
\frac{279565}{449079842304}\frac{\La_4^{14}}{v^{13/3}}\nonumber\\& & 
\hbox{}+\frac{5}{81}\frac{u^2}{v^{7/3}}\La_4^4+\frac{10}{81}
\frac{u^3}{v^{7/3}}\La_4^2+\ldots\\
\Omega_2&=&\omega_2\ln\frac{\La_4^3}{v}+\frac{5273}{1679616}
\frac{\La_4^8}{v^{7/3}}+
\frac{103}{2187}\frac{u}{v^{7/3}}\La_4^6+
\frac{9833323}{411411861504}\frac{\La_4^{14}}{v^{13/3}}\nonumber\\
& & \hbox{}+\frac{7}{27}\frac{u^2}{v^{7/3}}\La_4^4+\frac{11}{18}
\frac{u^3}{v^{7/3}}\La_4^2+\ldots
\eea

Having found the power series and  the logarithmic solutions it remains 
to calculate the period integrals to leading orders,
to determine the coefficients in  
$a_i= p_{i1}\,\om_1 + p_{i2} \,\om_2$ and 
$a_{Di}= q_{i1}\,\om_1 + q_{i2}\, \om_2 + q_{i3}\,\Om_1 + q_{i4} \,\Om_2$.

The first step in computing the period integrals $a_i=\int_{\alpha_i}\lam$ 
and $a_{Di}=\int_{\beta_i}\lam$ is to expand the six roots $e_i$
of the curve $y^2=\prod_i (x-e_i)$ around $v\to\infty$ with the result:
\bea\label{rootvinf}
e_1=\frac{\La_4}{3}+v^{1/3}+\frac{1}{9 v^{1/3}}\La_4^2+\ldots&{}
&e_4=-\frac{\La_4}{3}+v^{1/3}+\frac{1}{9 v^{1/3}}
\La_4^2+\ldots\nonumber\\
e_2=\frac{\La_4}{3}+\zeta v^{1/3}+\frac{\zeta^2}{9 v^{1/3}}\La_4^2
+\ldots&{}&e_5=-\frac{\La_4}{3}+\zeta v^{1/3}+
\frac{\zeta^2}{9 v^{1/3}}\La_4^2+\ldots\\
e_3=\frac{\La_4}{3}+\zeta^2 v^{1/3}+\frac{\zeta}{9 v^{1/3}}\La_4^2
+\ldots&{}&e_6=-\frac{\La_4}{3}+\zeta^2 v^{1/3}+\frac{\zeta}
{9 v^{1/3}}\La_4^2+\ldots\nonumber
\eea
where $\zeta=e^{2\pi i/3}$.

The pairs of roots which correspond to the basic cycles chosen in figure 1 are:
\be
\ba{llll}
\alpha_1\to (e_5,e_2)&{}&{}&\beta_1\to(e_5,e_4)\\
\alpha_2\to(e_6,e_3)&{}&{}&\beta_2\to(e_1,e_5)
\ea
\ee
In order to treat the poles of $\lam$ in the integrals separately 
we split the integration region by introducing an arbitrary parameter $\xi$. 
At the end of the calculation the period integrals will have to be independent of $\xi$. 
Inserting the meromorphic one form (\ref{la}), the integrals are 
of the form:
\be
I_i=\int_{e_i}^\xi dx\frac{(x^3+ux+2v)}{\{\prod_{j=1}^6(x-e_j)\}^{1/2}}
\ee
If we now introduce $\Del_1^{\pm}=\frac{1}{2}(e_1\pm e_4),\;
\Del_2^{\pm}=\frac{1}{2}(e_2\pm e_5),\;
\Del_3^{\pm}=\frac{1}{2}(e_3\pm e_6)$, 
$\Del_{i+3}^{\pm}=\pm\Del_i^{\pm}$, change variables such that  
$x=\rho\Del_i^{-}+\Del_i^+$ 
and use the expansion $(x^3+u x+2 v)=\sum_{k=0}^3\tilde{\eps}_k\rho^k$, we get:
\be\label{fint}
I_i=\int_1^{\nu_i}d\rho\frac{\sum_{k=0}^3\tilde{\eps}_k\rho^k}
{\sqrt{\rho^2-1}}\prod_{j\neq i\atop j\neq i-3}\bigg\{
\frac{1}{(\Del_j^+-\Del_i^+)(1-\rho\eps_j)} \Big(1+\frac{\sigma^2_j}
{2(1-\eps_j\rho)^2}+\ldots\Big)\bigg\}
\ee
where $\sigma_j=\frac{\Del_j^-}{\Del_j^+-\Del_i^+}$ and 
$\eps_j=\frac{\Del_i^-}{\Del_j^+-\Del_i^+}$ and 
$\nu_i=\frac{\xi-\Del_i^+}{\Del_i^-}$.
We can now express $I_i$ in terms of basic integrals 
$I^m_n=\int_1^{\nu}d\rho\frac{\rho^m}{\sqrt{\rho^2-1}(1-\eps\rho)^n}$.
By carefully doing the integrals in the complex plane, we get
the following result for the periods:
\bea
a_1=2\int_{e_2}^{e_5}\lam&=&2\Big(\zeta\omega_1
+\frac{1}{18}\zeta^2\omega_2\Big)\\
a_2=2\int_{e_6}^{e_3}\lam&=&2\Big(\zeta^2\omega_1
+\frac{1}{18}\zeta\omega_2\Big)\\
a_{D1}=2\int_{e_5}^{e_4}\lam&=&\frac{1}{i\pi 54}
\bigg\{\omega_2\Big(-3\ln(108)(2+\zeta)+ i\pi (17 \zeta+8)\Big)
\nonumber\\& & +\omega_1\Big(-54\ln(108)(1-\zeta)-i\pi 18 
(17 \zeta+9)+108 (1-\zeta)\Big)+2(2+\zeta)\Omega_2
\nonumber\\& & +36(1-\zeta)\Omega_1\bigg\}\\
a_{D2}=2\int_{e_1}^{e_3}\lam&=&\frac{1}{i\pi 54}
\bigg\{\omega_2\Big(3\ln(108)(\zeta-1)+i\pi(-13\zeta-9)\Big)\nonumber\\
& & +\omega_1\Big(54\ln(108)(-\zeta-2)+18(4 i \pi 
+13 i \pi\zeta+6\zeta+12)\Big)+2(1-\zeta)\Omega_2\nonumber\\
& & +36(\zeta+2)\Omega_1\bigg\}
\eea
Having expressed the periods in terms of power series and logarithmic solutions,
we can check the monodromies by taking $v \to e^{2\pi i}v$, resulting in
\be 
\tilde{M}_{v\to \infty}=\left(\ba{rrrr}-1&1&-5&0\\-1&0&-2&3\\0&0&0&1\\
0&0&-1&-1\ea\right)
\ee
This monodromy is $Sp(4,\ZZ)$-conjugate to (\ref{mvinfty}).

Similarly, we now construct solutions for the regime $u\to\infty$. 
There is one power series and
one logarithmic solution for the indices $(s,t)=(1/2,0)$:
\bea
\omega_1&=&\sqrt{u}+\frac{1}{16}\frac{\La_4^2}{\sqrt{u}}
-\frac{3}{1024}\frac{\La_4^4}{u^{3/2}}+
\frac{5}{16384}\frac{\La_4^6}{u^{5/2}}
 -\frac{175}{4194304}\frac{\La_4^8}{u^{7/2}}-\frac{3}{8}
\frac{v^2}{u^{5/2}}+\ldots\\
\Omega_1&=&\omega_1\ln\frac{\La_4^2}{u}-\frac{1}{8}\frac{\La_4^2}
{\sqrt{u}}+\frac{1}{1024}\frac{\La_4^4}{u^{3/2}}+
\frac{1}{49152}\frac{\La_4^6}{u^{5/2}}
-\frac{265}{25165824}\frac{\La_4^8}{u^{7/2}}-\frac{v^2}{u^{5/2}}
+\ldots
\eea
and another set of solutions for $(s,t)=(-1,1)$:
\bea
\omega_2&=&\frac{v}{u}+\frac{v^3}{u^4}+\frac{v^3}{u^5}\La_4^2+3
\frac{v^5}{u^7}+\frac{21}{2}\frac{v^5}{u^8}\La_4^2+
\frac{21}{4}\frac{v^5}{u^9}\La_4^4
+12\frac{v^7}{u^{10}}+90\frac{v^7}{u^{11}}\La_4^2+\ldots\\
\Omega_2&=&\omega_2\ln\Big(\frac{v^8\La_4^6}{u^{15}}\Big)
+\frac{1}{4}\frac{v}{u^2}\La_4^2-
\frac{1}{32}\frac{v}{u^3}\La_4^4+\frac{1}{192}\frac{v}{u^4}\La_4^6
 -\frac{1}{1024}\frac{v}{u^5}\La_4^8+\frac{119}{6}\frac{v^3}{u^4}
+\frac{1}{5120}\frac{v}{u^6}\La_4^{10}+\ldots
\eea

The roots of the curve expand in the region $u\to\infty$ as:
\bea
e_1=-\frac{v}{u}+\frac{v^2}{ u^{3}}\La_4+\ldots&{}&e_4=-\frac{v}{u}
-\frac{v^2}{u^{3}}\La_4+\ldots\nonumber\\
e_2=-\frac{\La_4}{2}+\sqrt{u}+\ldots&{}&e_5=\frac{\La_4}{2}+\sqrt{u}
+\ldots\\
e_3=-\frac{\La_4}{2}-\sqrt{u}+\ldots&{}&e_6=\frac{\La_4}{2}-\sqrt{u}
+\ldots\nonumber
\eea

The integrals for the periods are calculated in the same manner as before.
\bea
a_1&=&2\int_{e_5}^{e_2}\lam=2\Big(-\frac{1}{2}\om_2-\om_1\Big)\\
a_2&=&2\int_{e_6}^{e_3}\lam=2\Big(\frac{1}{2}\om_2-\om_1\Big)\\
a_{D1}&=&2\int_{e_2}^{e_1}\lam=\frac{2}{i\pi}
\Big(\om_2\Big(2+\frac{3\ln2}{2}\Big)+\om_1(-1+3\ln2)
-\frac{1}{4}\Om_2-\frac{1}{2}
\Om_1\Big)\\
a_{D2}&=&2\int_{e_4}^{e_6}\lam=\frac{2}{i\pi}
\Big(\om_2\Big(-2-\frac{3\ln2}{2}\Big)+\om_1(-1+3\ln2)+\frac{1}{4}\Om_2-
\frac{1}{2}\Om_1\Big)
\eea
The resulting monodromy matrix of the periods in the 
regime $u\to\infty$ conjugate to $M_{u\to\infty}$ (\ref{muinfty}) is:
\be
\tilde{M}_{u\to\infty}=\left(\ba{rrrr}0&-1&-7&8\\-1&0&8&-7\\0&0&0&
-1\\0&0&-1&0\ea\right)
\ee


\subsection{Prepotential}

In the previous section we found two power series and two 
logarithmic solutions of the Picard-Fuchs operators
for the semiclassical regions $u\to \infty$ and $v\to \infty$ 
and determined the periods $a_i(u,v)$ 
and $a_{Di}(u,v)$.
In this section we calculate the prepotential $\cF(a_1,a_2)$ 
by integration.
 
Since the periods $a_i(u,v)$ are expressed in terms of the Casimirs 
$u$ and $v$, the prepotential
is readily obtained by integrating the following two equations:
\bea
\frac{\p\cF}{\p u}&=&a_{Di}(u,v)\frac{\p a_i(u,v)}{\p u}\label{fffu}\\
 \frac{\p\cF}{\p v}&=&a_{Di}(u,v)\frac{\p a_i(u,v)}{\p v}\label{fffv}
\eea
The integrability condition $\frac{\p a_{Di}}{\p v}\frac{\p a_i}{\p u}
-\frac{\p a_{Di}}{\p u}\frac{\p a_i}{\p v}=0$
can be used as a check for the period integrals $a_i$ and $a_{Di}$.
Integrating (\ref{fffu},\ref{fffv}) yields the prepotential 
$\cF(u,v)=\cF_{\rm class}(u,v)+\cF_{\rm 1-loop}(u,v)+\cF_{\rm inst}(u,v)$, 
which contains  power series in $u$, $v$ and $\La_4$, as well as logarithmic terms.
Our aim is to express the prepotential $\cF$ in terms of the periods $a_i$. 
For that we introduce the central charges 
$Z_i=\la\vec\alpha_i,\vec a\ra$ with $\vec\alpha_i\in\Del_+(A_2)$ 
and $\vec a=(a_1,a_2)^T$:
\bea
Z_1&=&2 a_1-a_2\nonumber\\
Z_2&=&2 a_2-a_1\\
Z_3&=&Z_1+Z_2\nonumber
\eea
In these variables the prepotential is a homogenous function of degree two.
The classical prepotential $\cF_{\rm class}$ is proportional to 
$\sum_{i=1}^3 Z_i^2$ and the one 
loop part contains logarithms of $Z_i$ multiplied with homogenous 
polynomials in $Z_i$ 
of degree two. The proportionality constants can be found by 
matching the expressions as functions of $u$ and $v$ 
against $\cF(u,v)$. In this way one obtains simultaneously 
 $\cF_{\rm class}$ and $\cF_{\rm 1-loop}$ as functions of $a_1$ and $a_2$. 
For $u\to \infty$ we get:
\bea
\cF_{\rm class}&=&\frac{1}{4i\pi}\sum_{i=1}^3 Z_i^2\\
\cF_{\rm 1-loop}&=&-\frac{1}{4 i\pi}\sum_{i=1}^3 Z_i^2 \ln\left(\frac{Z_i}{\La_4}\right)^2+ \frac{1}{2i\pi} \Big\{\Big(\frac{Z_1-Z_2}{3}\Big)^2\ln\left(\frac{Z_1-Z_2}
{3\La_4}\right)^2+\nonumber\\
& & \hbox{}+\Big( \frac{Z_2+Z_3}{3}\Big)^2\ln\left(\frac{Z_2+Z_3}
{3\La_4}\right)^2 +\Big(\frac{Z_1+Z_3}{3}\Big)\ln\left(\frac{Z_1+Z_3}
{3\La_4}\right)^2\Big\}\nonumber
\eea
For $v\to \infty$ we find the same result up to an overall minus sign.
This observation holds for all results in this and the following section.

After subtracting the classical and the one loop part 
from the prepotential a  power series in $\La_4^2$ 
remains which gives the instanton contributions. The individual 
contributions can  be summed up in terms of $Z_i$. Here we give the result for 
the one and two instanton corrections, cf. (\ref{1lp}):
\bea\label{f1nf4}
\cF_1&=&\frac{i}{3\pi} \Big(1-\frac{u_0^3}{\Del_0}\Big)\\
\cF_2&=&-\frac{i}{\pi}\Big(\frac{1}{6}\frac{u_0^2}{\Del_0}
-\frac{7}{6}\frac{u_0^5}{\Del_0^2}+\frac{5}{2}\frac{u_0^8}{\Del_0^3}\Big)
\eea
where $u_0=a_1^2 + a_2^2 -a_1a_2 = \frac 16 \sum_i Z_i^2$ and $\Del_0=\prod_i Z_i^2$.

\section{$N_f=2$}

\subsection{Monodromies}

Proceeding similarly to the case $N_f=4$ we determine 
the  monodromies for $N_f=2$ by fixing $v=1$, $u=-2$ as a 
base point in the moduli space ${\cal M}_{N_f=2}$
and looping around the zero loci of $\Del_{N_f=2}$. Tracing 
the motion of the roots of the curve
in the $x$-plane leads to the vanishing cycles, from which 
we determine the monodromies.
\begin{figure}[htp]
\begin{center}
\input{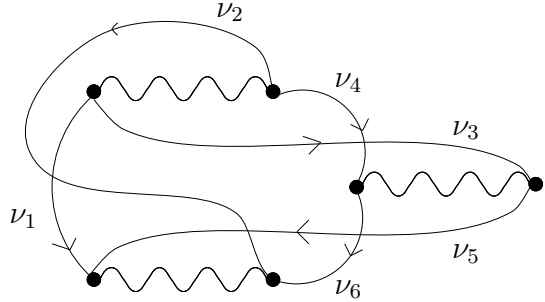}
\end{center}
\caption{Vanishing cycles for $N_f=2$}
\end{figure}

If we take the basic cycles to be the same as in the case 
$N_f=4$, we get the following vanishing cycles:
\bea
\nu_1&=&(1,1,1,0)\nonumber\\
\nu_2&=&(1,1,2,1)\nonumber\\
\nu_3&=&(1,0,3,-1)\\
\nu_4&=&(1,0,1,0)\nonumber\\
\nu_5&=&(0,1,0,-1)\nonumber\\
\nu_6&=&(0,1,-1,1)\nonumber
\eea

In the case $N_f=2$ the structure of the moduli space can easily 
be read off from (\ref{disc2}), which reads for $\La_2=1$
\be \Del_{N_f=2}= v^2 (4(u+1)^3 - 27 v^2 )(4(u-1)^3 - 27 v^2) \ee
There are two cusps at $u=\pm 1$, $v=0$ at each of which 6 branches meet. 
In addition there are 4 nodes at the points $u=i / \sqrt 3$, 
$v=\pm 9(1 + i)/(8 \sqrt[4]{3})$
and $u=- i / \sqrt{3}$, $v=\pm 9(1 - i)/(8 \sqrt[4]{3})$.

The semiclassical monodromy in the region $v\to\infty$ is:
 \be\label{mmvi}
{M}_{v\to\infty}=(M_1\cdot M_7\cdot M'_6\cdot M'_5)^{-1}=
\left(\ba{rrrr}-1&-1&-4&0\\1&0&6&-2\\0&0&0&-1\\0&0&1&-1\ea\right)
\ee
where $M_7$ comes from $\nu_7=(1,1,0,-1)$ and $M'_5$ and $M'_6$
coincide with $M_5$ and $M_6$.
In the region $u\to \infty$ we get:
 \be\label{mmui}
{M}_{u\to\infty}=M_6\cdot M_5\cdot M_2\cdot M_1\cdot M_3\cdot M_4 =
\left(\ba{rrrr}0&-1&-7&8\\-1&0&8&-7\\0&0&0&-1\\0&0&-1&0\ea\right)
\ee


\subsection{Solution  in the semiclassical regions}

The procedure is identical to the one in section 3.2, we will therefore be brief.
Solving the recursion relation for $v\to\infty$ gives one 
power series and one logarithmic
solution associated with the indices $s=0$ and $t=\frac{1}{3}$:
\bea
\om_1&=& v^{1/3}-\frac{1}{54}\frac{u}{v^{5/3}}\La_2^4
-\frac{5}{26244}\frac{\La_2^{12}}{v^{11/3}}-
\frac{1}{81}\frac{u^3}{v^{5/3}}-\frac{5}{1458}
\frac{u^2}{v^{11/3}}\La_2^8-\frac{10}{2187}\frac{u^4}{v^{11/3}}\La_2^4
+\ldots\\
\Om_1&=&\om_1\ln\frac{\La_2^3}{v}+\frac{1}{36}\frac{u}{v^{5/3}}\La_2^4+\frac{7}{26244}\frac{\La_2^{12}}{v^{11/3}}+
\frac{1}{324}
\frac{u^2}{v^{11/3}}\La_2^8+\ldots
\eea
A second solution can be found for $s=0$ and $t=-\frac{1}{3}$:
\bea
\om_2&=&\frac{u}{v^{1/3}}+\frac{1}{216}\frac{\La_2^8}{v^{7/3}}
+\frac{1}{27}\frac{u^2}{v^{7/3}}\La_2^4+
\frac{175}{104976}\frac{u}{v^{13/3}}\La_2^{12}
+\frac{91}{5668704}\frac{1}{v^{19/3}}\La_2^{20}
+\frac{1}{81}\frac{u^4}{v^{7/3}}\nonumber\\
& & +\frac{175}{17496}\frac{u^3}{v^{13/3}}\La_2^{12}
+\frac{35}{4374}\frac{u^5}{v^{13/3}}\La_2^4+\ldots\\
\Om_2&=&\om_2\ln\frac{\La_2^3}{v}+\frac{1}{96}\frac{\La_2^8}{v^{7/3}}
+\frac{1}{9}\frac{u^2}{v^{7/3}}\La_2^4+
\frac{1945}{419904}\frac{u}{v^{13/3}}\La_2^{12}
+\frac{1}{18}\frac{u^4}{v^{7/3}}+\nonumber\\
& & \frac{85}{2592}\frac{u^3}{v^{13/3}}\La_2^8+\ldots
\eea

 We expand the roots in the semiclassical region $v\to\infty$, 
that are needed for calculating the integrals:
\bea
e_1=v^{1/3}+\frac{\La_2+u}{ 3 v^{1/3}}+\ldots&{}&
e_4=v^{1/3}-\frac{\La_2-u}{3 v^{1/3}}+\ldots\nonumber\\
e_2=\zeta v^{1/3}+\frac{\zeta^2(\La_2+u)}{3 v^{1/3}} 
+\ldots&{}&e_5=\zeta v^{1/3}-\frac{\zeta^2(\La_2-u)}{3 v^{1/3}}+\ldots\\
e_3=\zeta^2 v^{1/3}+\frac{\zeta(\La_2+u)}{3 v^{1/3}}  
+\ldots&{}&e_6=\zeta^2 v^{1/3}-\frac{\zeta(\La_2-u)}{3 v^{1/3}}+\ldots
\nonumber
\eea

The period integrals are calculated in a way parallel to the case $N_f=4$ with the result:
\bea
a_1=2\int_{e_5}^{e_2}\lam &=&2\Big(\om_1\zeta+\frac{\om_2}{3}\zeta^2\Big)\\
a_2=2\int_{e_6}^{e_3}\lam &=&2\Big(-\om_1\zeta^2-\frac{\om_2}{3}\zeta\Big)\\
a_{D1}=2\int_{e_2}^{e_4}\lam &=&\frac{1}{9 i \pi}
\Big\{\om_1\Big(-57 i \pi \zeta-27 i \pi+36 (1-\zeta)
+9 \ln (108) (\zeta-1)\Big)
+ \nonumber\\
& & \om_2\Big(-3\ln 108(\zeta+2)+19 i\pi\zeta+10 i\pi\Big)
+12(1-\zeta)\Om_1+\nonumber\\
& & (\zeta+2)4 \Om_2\Big\}\\
a_{D2}=2\int_{e_1}^{e_6}\lam &=&-\frac{\zeta}{9 i \pi}
\Big\{\om_1\Big(\ln (108) 9 (1+2 \zeta)-6 i\pi(\zeta^2+2 \zeta)
+21 i\pi-36(1+2 \zeta)\Big)+\nonumber\\
& & \om_2\Big(3\ln 108 (\zeta+2)+2 i\pi\zeta^2
+11 i\pi \zeta\Big)-12 \Om_1(1+2 \zeta)-\nonumber\\
& & 4\Om_2(2+\zeta)\Big\}
\eea
They undergo the semiclassical monodromy 
 \be
M_{v\to\infty}=\left(\ba{rrrr}-1&-1&-6&0\\1&0&4&2\\0&0&0&-1\\0&0&1&-1\ea\right)
\ee
which is conjugated to (\ref{mmvi}).

The region $u\to\infty$ gives rise to a set of solutions with 
indices $(s=\frac{1}{2},t=0)$ and $(s=-1,t=1)$.
For $s=\frac{1}{2}$ and $t=0$ we get one power series and one 
logarithmic solution:
\bea
\om_1&=&\sqrt{u}-\frac{1}{16}\frac{\La_2^4}{u^{3/2}}
-\frac{15}{1024}\frac{\La_2^8}{u^{7/2}}
-\frac{3}{8}\frac{v^2}{u^{5/2}}-\frac{105}{16384}
\frac{\La_2^{12}}{u^{11/2}}-\frac{105}{128}\frac{v^2}{u^{9/2}}\La_2^4
-\frac{15015}{4194304}\frac{\La_2^{16}}{u^{15/2}}+\ldots\\
\Om_1&=&\om_1\ln\frac{\La_2^2}{u}+\frac{1}{16}
\frac{\La_2^4}{u^{3/2}}+\frac{13}{2048}\frac{\La_2^8}{u^{7/2}}-
\frac{1}{4}\frac{v^2}{u^{5/2}}
+\frac{163}{98304}\frac{\La_2^{12}}{u^{11/2}}
-\frac{37}{128}\frac{v^2}{u^{9/2}}\La_2^4+\ldots
\eea
and  a second set of solutions for $s=-1$ and $t=1$:
\bea
\om_2&=&\frac{v}{u}+\frac{1}{2}\frac{v}{u^3}\La_2^4
+\frac{3}{8}\frac{v}{u^5}\La_2^8+\frac{v^3}{u^4}
+\frac{5}{16}\frac{v}{u^7}\La_2^{12}+5\frac{v^3}{u^6}\La_2^{4}
+\frac{35}{128}\frac{v}{u^9}\La_2^{16}+\ldots\\
\Om_2&=&\om_2\ln\frac{v\La_2^3}{u^3}+\frac{3}{2}
\frac{v}{u^3}\La_2^4+\frac{3}{2}\frac{v}{u^5}\La_2^8
+\frac{25}{6}\frac{v^3}{u^4}+\frac{23}{16}\frac{v}{u^7}
\La_2^{12}+\frac{241}{12}\frac{v^3}{u^6}\La_2^4+\ldots
\eea

For the integration we  expand the roots of the curve in 
the limit of large $u$:
\bea
e_1=-\frac{v}{u}-3\frac{v}{ u^{2}}\La_2^2+\ldots&{}
&e_4=-\frac{v}{u}+3\frac{v}{u^{2}}\La_2^2+\ldots\nonumber\\
e_2=\sqrt{u}+\frac{1}{2}(-\frac{\La_2^2}{\sqrt{u}}
+\frac{v}{u})+\ldots&{}&e_5=\sqrt{u}+\frac{1}{2}
(\frac{\La_2^2}{\sqrt{u}}+\frac{v}{u})+\ldots\\
e_3=-\sqrt{u}+\frac{1}{2}(\frac{\La_2^2}{\sqrt{u}}
+\frac{v}{u})+\ldots&{}&e_6=-\sqrt{u}+
\frac{1}{2}(-\frac{\La_2^2}{\sqrt{u}}+\frac{v}{u})+\ldots\nonumber
\eea

For the period integrals we find:
\bea
a_1&=&2\int_{e_5}^{e_2}\lam=2\Big(-\om_1-\frac{1}{2}\om_2\Big)\\
a_2&=&2\int_{e_3}^{e_6}\lam=2\Big(-\om_1+\frac{1}{2}\om_2\Big)\\
a_{D1}&=&2\int_{e_2}^{e_4}\lam=\frac{2}{i\pi}\Big(\om_1(-2+3\ln2)
+\om_2\Big(1+\frac{3}{2}\ln2\Big)-
\Om_1-\Om_2\Big)\\
a_{D2}&=&2\int_{e_1}^{e_3}\lam=\frac{2}{i\pi}\Big(\om_1(-2+3\ln 2)
+\om_2\Big(-1 -\frac{3\ln2}{2}\Big)-
\Om_1+\Om_2\Big)
\eea
which leads to the following monodromy matrix:
\be
\tilde{M}_{u\to\infty}=\left(\ba{rrrr}0&-1&-7&9\\-1&0&5&-3\\0&0&0&
-1\\0&0&-1&0\ea\right)
\ee
This is conjugated to (\ref{mmui}).


\subsection{Prepotential}
As in the $N_f=4$ case, the instanton corrections to the prepotential are 
obtained by subtracting 
$\cF_{\rm class}$ and $\cF_{\rm 1-loop}$ from $\cF(u,v)$.
For both weak-coupling regions we find:
\bea
\cF_{\rm class}&=&\frac{1}{2i\pi}\sum_{i=1}^3 Z_i^2\Big(1+\frac{\ln 2}{3}\Big)\\
\cF_{\rm 1-loop}&=&-\frac{1}{4 i\pi}\sum_{i=1}^3 Z_i^2 \ln\left(\frac{Z_i}{\La_2}\right)^2+
\frac{1}{4 i \pi}\Big\{\Big(\frac{Z_1-Z_2}{3}\Big)^2\ln\left(\frac{Z_1-Z_2}{3\La_2}\right)^2+\\
& &  \Big(\frac{Z_2+Z_3}{3}\Big)^2\ln\left(\frac{Z_2+Z_3}
{3\La_2}\right)^2+ \Big(\frac{Z_1+Z_3}{3}\Big)^2\ln\left(\frac{Z_1+Z_3}{3\La_2}\right)^2\Big\}\nonumber
\eea

The coefficients of $\La_2^4$ and $\La_2^8$ in the instanton series sum up to:
\bea\label{f1nf2}
\cF_1&=&-\frac{4 i}{\pi} \frac{u_0^2}{\Del_0}\\
\cF_2&=&-\frac{i}{\pi}\Big(\frac{8}{\Del_0}-112\frac{u_0^3}{\Del_0^2}
+360 \frac{u_0^6}{\Del_0^3}\Big)
\eea
with $u_0$ and $\Del_0$ as before.


\section{Instanton Corrections by an Alternative Method}

So far we have seen how to determine the prepotential from the 
periods $a_i$ and $a_{Di}$. 
This requires knowledge of all four solutions of the Picard-Fuchs equations.
However, our ansatz for the logarithmic solutions does not work for all $N_f$.
We now present an alternative way to derive the prepotential which requires only knowledge of the power series solutions of the Picard-Fuchs equations.\footnote{This 
method can also be applied to other patches of the moduli space.}
We will apply this method to $N_f=1,3$ and 5 in the semiclassical region $u\to\infty$
and check our previous results for $N_f=2$ and 4.

Classically, i.e. for $\La_{N_f} \to 0$, the Casimirs are given by:
\bea\label{u0v0}
u_0&=&\frac{1}{6}\sum_{i=1}^3 Z_i^2=a_1^2+a_2^2-a_1 a_2\nonumber\\
v_0&=&a_1 a_2 (a_1-a_2)
\eea
Inverting the above equations we get:
\bea
a_1&=&p_{+}+p_{-}\nonumber\\
a_2&=&-\zeta p_{+}-\zeta^2 p_{-}
\eea
where $p_{\pm}=\Big(\frac{v_0}{2}\pm\frac 12\sqrt{v_0^2-\frac{4}{27} u_0^3}\Big)^{1/3}$. 
On the other hand, evaluating $a_i$ for $u_0\to\infty$ we get 
$a_1\sim\sqrt{u_0}+\hf \frac{v_0}{u_0}$ and $a_2\sim\sqrt{u_0}-\hf\frac{v_0}{u_0}$.

Since we know that in this region the Picard-Fuchs equations 
give two power series solutions $\om_1$ and $\om_2$
with asymptotic behaviour
$\om_1=\sqrt{u}+\ldots$ and $\om_2=\frac{v}{u}+\ldots$, 
the periods must be of the form
$a_1=-2(\om_1+\hf \om_2)$ and $a_2=-2(\om_1-\hf \om_2)$, taking the normalization 
from our previous conventions for the period integrals. 

Now we use a relation between $u$ and $\cF$ derived in \cite{m,sty,ey} 
to obtain the instanton corrections of the prepotential:
\bea\label{uinst}
u(Z)&=&\frac{4 i \pi}{(2N_c-N_f)} \Big(\cF-\hf Z_j\frac{\p\cF}{\p Z_j}\Big)
\nonumber\\
&=&\frac{1}{6}\sum_{i=1}^3 Z_i^2+2 i \pi \sum_{n=1}^{\infty} 
\cF_n(Z) n \La_{N_f}^{(2 N_c-N_f) n}
\eea
 $u$ itself has a power series expansion in $\La_{N_f}$, namely 
$u(Z)=u_0+\sum_{n=1}^{\infty}\cG_n(Z)\La_{N_f}^{(2 N_c-N_f)n}$ 
with $u_0$ defined in (\ref{u0v0}).
Summing the series for $u$ by rewriting it in the variables 
$Z_i$ yields the instanton corrections to the prepotential.

For $N_f=1,\ldots,5$ we determine the two power series $\om_1$ 
and $\om_2$ again by using the Picard-Fuchs equations 
and find for $u$
\bea
N_f=1:\hspace{1cm} u&=&u_0- 18 \frac{v_0}{\Del_0}\La_1^5
-\Big(1008 \frac{u_0^2}{\Del_0^2}-
4320\frac{u_0^5}{\Del_0^3}\Big)\La_1^{10}+\cO(\La_1^{15})\nonumber\\
N_f=2:\hspace{1cm} u&=& u_0+ 8 \frac{u_0^2}{\Del_0}\La_2^4
+ 4 \Big(\frac{8}{\Del_0}-112 \frac{u_0^3}{\Del_0^2}
+360 \frac{u_0^6}{\Del_0^3}\Big)\La_2^8+\cO(\La_2^{12})\nonumber\\
N_f=3:\hspace{1cm} u&=&u_0-\frac{3}{2}\frac{u_0 v_0}{\Del_0}\La_3^3
-\Big(-\frac{2}{3}\frac{u_0}{\Del_0}+
\frac{35}{3}\frac{u_0^4}{\Del_0^2}-30\frac{u_0^7}{\Del_0^3}\Big)\La_3^6
+\cO(\La_3^9)\\
N_f=4:\hspace{1cm} u&=&u_0-\frac{2}{3}
\Big(1-\frac{u_0^3}{\Del_0}\Big)\La_4^2+4 \Big(\frac{1}{6}\frac{u_0^2}{\Del_0}
-\frac{7}{6}\frac{u_0^5}{\Del_0^2}+\frac{5}{2}\frac{u_0^8}{\Del_0^3}\Big)\La_4^4
+\cO(\La_4^6)\nonumber\\
N_f=5:\hspace{1cm} u&=&u_0- \frac{3}{2}\frac{u_0^2 v_0}{\Del_0}\La_5
-\Big(-\frac{10}{3}-2 \frac{u_0^3}{\Del_0}
+\frac{49}{3}\frac{u_0^6}{\Del_0^2}-30\frac{u_0^9}{\Del_0^3}\Big)\La_5^2
+\cO(\La_5^3)\nonumber
\eea
From these equations and (\ref{uinst}) we can read off the one 
and two instanton corrections for all $N_f$.
For the two cases $N_f=2,4$ calculated at length above we can 
compare the results. For example for $N_f=2$ 
we have $\cF_1=\frac{\cG_1}{2 i\pi}=-\frac{4 i}{\pi}\frac{u_0^2}{\Del_0}$ and 
$\cF_2=\frac{\cG_2}{4 i\pi}=-\frac{i}{\pi}\Big(\frac{8}{\Del_0}
-112\frac{u_0^3}{\Del_0^2}+360 \frac{u_0^6}{\Del_0^3}\Big)$ 
which agrees with (\ref{f1nf2}).
For $N_f=4$ we find agreement with the results  (\ref{f1nf4}).

For the one and two instanton contributions for $N_f=1,3,5$ we find:
\bea
N_f=1:\hspace{1cm} \cF_1&=&9\frac{i }{\pi}\frac{v_0}{\Del_0}\\
\cF_2&=&\frac{i}{\pi}\Big(252\frac{u_0^2}{\Del_0^2}
-1080\frac{u_0^5}{\Del_0^3}\Big) \nonumber\\
N_f=3:\hspace{1cm} \cF_1&=&\frac 34\frac{i }{\pi}\frac{u_0 v_0}{\Del_0}\\
\cF_2&=&\frac{i}{\pi}\Big(-\frac{1}{6}\frac{u_0}{\Del_0}
+\frac{35}{12}\frac{u_0^4}{\Del_0^2}-
\frac{15}{2}\frac{u_0^7}{\Del_0^3}\Big)\nonumber\\  
N_f=5:\hspace{1cm} \cF_1&=&\frac 34\frac{i }{\pi}\frac{u_0^2 v_0}{\Del_0}\\
\cF_2&=&\frac{i}{\pi}\Big(-\frac{5}{6}-\frac{u_0^3}{2\Del_0}
+\frac{49}{12}\frac{u_0^6}{\Del_0^2}-
\frac{15}{2}\frac{u_0^9}{\Del_0^3}\Big)
\eea

With both methods we can easily calculate higher order instanton corrections as well.
It would be nice to verify these results by explicit instanton calculation.


\end{document}